\documentclass[prd,aps,letterpaper,nofootinbib,notitlepage,twocolumn,10pt]{revtex4-2}
\usepackage[a4paper, hdivide={1.91cm,,1.165cm}, vdivide={1.83cm,,2.1cm}]{geometry}

\usepackage{amsmath,amssymb}
\usepackage{graphicx,multirow}
\usepackage{units}
\usepackage{supertabular}
\usepackage{longtable}
\usepackage[hyperfootnotes=false]{hyperref}

\def \L {\mathcal{L}} 

\def \epsilon {\varepsilon} 

\def \vec#1{{\boldsymbol{#1}}}
\newcommand{\hc}{\ensuremath{\text{h.c.}}}
\newcommand{\BR}{\ensuremath{\text{BR}}}
\newcommand{\dd}{\ensuremath{\text{d}}}
\newcommand{\ii}{\ensuremath{\text{i}}}

\newcommand{\Element}[3]{${}^{#3}_{#2}\text{#1}$}

\allowdisplaybreaks

\begin{document}

\title{Isotope dependence of muon-to-electron conversion}

\author{Julian Heeck}
\email{heeck@virginia.edu}
\affiliation{Department of Physics, University of Virginia,
Charlottesville, Virginia 22904-4714, USA}

\author{Robert Szafron}
\email{rszafron@bnl.gov}
\affiliation{Department of Physics, Brookhaven National Laboratory, Upton, N.Y., 11973, USA}

\author{Yuichi Uesaka}
\email{uesaka@ip.kyusan-u.ac.jp}
\affiliation{Faculty of Science and Engineering, Kyushu Sangyo University,
2-3-1 Matsukadai, Higashi-ku, Fukuoka 813-8503, Japan}

\begin{abstract}
The lepton-flavor-violating conversion of a muon into an electron in the field of a nucleus is one of the most sensitive probes of physics beyond the Standard Model and the experiments Mu2e, COMET, and DeeMe will explore uncharted terrain in the near future. An observation of this $\mu^-\to e^-$ conversion process opens up the possibility to distinguish the underlying operator or new-physics model by exploiting the target-nucleus dependence of the conversion rate. To facilitate the choice of optimal targets we provide $\mu^-\to e^-$ conversion rates for all stable isotopes and estimate nuclear-structure uncertainties. Our results enable studies of mixed or enriched target materials that are particularly promising for distinguishing scenarios in which the muon converts either on protons or neutrons.
\end{abstract}

\maketitle


\section{Introduction}

Charged lepton flavor violation (LFV) is highly suppressed in the Standard Model and therefore any observation of it would be a strong signal for the  physics beyond the Standard Model~\cite{Kuno:1999jp,Bernstein:2013hba}. 
Searches for rare \emph{muon} decays are among the most promising due to their relatively long lifetime and availability of intense muon sources. 
One of the most sensitive search avenues is the coherent $\mu^-\to e^-$ conversion in the field of a nucleus, $\mu^- +(A,Z) \to e^- +(A,Z)$. It is particularly sensitive to effective LFV operators involving quarks~\cite{Crivellin:2017rmk,Davidson:2018kud}.
New experiments DeeMe~\cite{Natori:2014yba,Teshima:2019orf}, COMET~\cite{Adamov:2018vin,Moritsu:2021fns}, and Mu2e~\cite{Bartoszek:2014mya,Yucel:2021vir} are under development and they are expected to improve the existing limits on the conversion rate by several orders of magnitude. 
While the conversion searches target New Physics discovery, it is possible to differentiate between different LFV effective operators using complementary target nuclei in the event of a positive signal observation using the $\mu^-\to e^-$ conversion rate dependence on the target nucleus (notably its charge $Z$),\cite{Kitano:2002mt,Cirigliano:2009bz,Bartolotta:2017mff,Davidson:2017nrp,Davidson:2018kud,Davidson:2020hkf}.
The experiment DeeMe at J-PARC will use graphite or silicon carbide targets; Fermilab's Mu2e and J-PARC's COMET plan to search for the conversion in aluminium during their first phases. Previous experiments set limits using various other targets: copper~\cite{Bryman:1972rf}, sulfur~\cite{Badertscher:1980bt}, lead~\cite{SINDRUMII:1996fti}, titanium~\cite{SINDRUMII:1993gxf}, and gold~\cite{SINDRUMII:2006dvw}.

Muon-to-electron conversion depends on the structure of the nucleus, which makes the decay-rate calculations more complicated than those of the competing LFV processes $\mu\to e\gamma$ and $\mu\to 3e$. Increasingly advanced calculations have been performed in Refs.~\cite{Weinberg:1959zz,Marciano:1977cj,Shanker:1979ap,Czarnecki:1998iz,Crivellin:2014cta}.
Ref.~\cite{Kitano:2002mt} has provided a detailed study of the conversion rates in 55 nuclei that have been used extensively in the literature, notably to evaluate which targets are most useful in the event of an observation to extract the underlying new-physics operator~\cite{Cirigliano:2009bz,Davidson:2018kud,Davidson:2020hkf}.
The nuclear data underlying Ref.~\cite{Kitano:2002mt} has changed little over the last two decades and does not warrant a dedicated update. This article instead aims to extend Ref.~\cite{Kitano:2002mt} to cover all stable isotopes with natural abundance above $1\%$. This then allows for muon-to-electron studies on mixed or enriched target materials. As we will see below, this can significantly improve the complementarity of different targets. Our results also provide an independent cross-check of earlier  evaluations.

The energy of the electron produced in the $\mu^-\to e^-$ conversion is given by 
\begin{equation}
    E_e^{\rm conv} = m_\mu - E_b - E_{\rm recoil},
\end{equation}
where $m_\mu=\unit[105.66]{MeV}$ is the muon mass. The binding energy $E_b$, which in the non-relativistic approximation for a point-like nucleus is given by  $E_b \simeq  \alpha^2 Z^2 m_\mu/2 $, depends strongly on the nuclear target charge $Z$, while the dependence on the neutron number enters only though finite-nuclear size corrections~\cite{Heeck:2021adh}. For the nuclear recoil, $ E_{\rm recoil} = (m_\mu-E_b)^2/(2 m_N)$, the isotope dependence enters additionally through the nucleus mass $m_N$. 
Overall, the dependence of $E_e^{\rm conv}$ on the neutron number is small, so the experimental signature of a mono-energetic electron is largely independent of the isotope composition~\cite{Heeck:2021adh}.\footnote{Beyond the leading approximation, the electrons are no longer mono-energetic. The energy is smeared by various effects, notably by higher-order electromagnetic corrections~\cite{Szafron:2017guu}. The effect is comparable in size with the corrections that suppress the background due to the standard lepton number conserving decay of a bound muon into an electron and a neutrino pair~\cite{Szafron:2015kja, Szafron:2016cbv}.}
 This conclusion does not hold for the  $\mu^-\to e^-$ conversion \emph{rate}, which shows a significant isotope dependence as we will show below.

\section{Muon-to-electron conversion}

Since $\mu\to e $ conversion is a low-energy process, where typical energy scales involved are of the order of the muon mass, it can be described to leading order in chiral perturbation theory by an effective Lagrangian connecting muons and electrons via dipole and two-nucleon operators~\cite{Davidson:2018kud}:
\begin{align}
\L_{\mu e} &= -\frac{4 G_F}{\sqrt{2}}  \sum_{X = L,R}\left[ m_\mu C_{D,X} \overline{e} \sigma^{\alpha\beta} P_X \mu \,F_{\alpha\beta} \right.\nonumber\\
&+\sum_{N=p,n}\left( C_{S, X}^{(NN)} \overline{e} P_X \mu \, \overline{N} N
+ C_{P, X}^{(NN)} \overline{e} P_X \mu \,  \overline{N} \gamma_5 N \right.\nonumber\\
&+ C_{V, X}^{(NN)} \overline{e} \gamma^\alpha P_X \mu \,  \overline{N} \gamma_\alpha N
+ C_{A, X}^{(NN)} \overline{e} \gamma^\alpha P_X \mu \,  \overline{N} \gamma_\alpha \gamma_5 N\nonumber\\
&+ C_{Der, X}^{(NN)} \overline{e} \gamma^\alpha P_X \mu \,  (\overline{N} \overleftrightarrow{\partial}_\alpha\ii \gamma_5 N)\nonumber\\
&+ \left.\left.C_{T, X}^{(NN)} \overline{e} \sigma^{\alpha\beta} P_X \mu \,  \overline{N} \sigma_{\alpha\beta} N\right)
\right] + \hc ,
\label{eq:lagrangian}
\end{align}
where $P_{L,R}$ are chiral projection operators and the $C$ are dimensionless Wilson coefficients at the experimental energy scale $\sim m_\mu$.
We expect spin-\emph{independent} $\mu\to e$ conversion to dominate due to coherent enhancement.\footnote{For a recent computation of the conversion rate for $^{208}\rm Pb$ with a breakdown into coherent, and incoherent spin-dependent and spin-independent contributions, see Ref.~\cite{Civitarese:2019cds}.} 
This is an assumption that does not hold true in all possible models~\cite{Cirigliano:2017azj,Davidson:2017nrp} but will be employed from here on out.
For spin-\emph{independent} scattering, only a subset of Wilson coefficients contribute, leading to the $\mu\to e$ conversion rate, conventionally normalized relative to the muon capture rate $\Gamma_\text{capture}$~\cite{Kitano:2002mt,Davidson:2018kud},
\begin{align}
 & \BR_\text{SI}(\mu A\to e A) = \frac{32 G_F^2}{\Gamma_\text{capture}} \left[  
 \left| C^{pp}_{V,R} V^{(p)} + 
C^{pp'}_{S,L} S^{(p)} \right.\right.\nonumber\\
&\, \left.\left.+ 
C^{nn}_{V,R} V^{(n)} + 
C^{nn'}_{S,L} S^{(n)} + 
C_{D,L} \frac{D}{4}   \right|^2 + \{ L\leftrightarrow R\}\right] .
\label{eq:BR}
\end{align}
Here, the primed coefficients are defined as $C^{N N'}_{S,X} \equiv C^{N N}_{S,X} + \frac{2 m_\mu}{m_N} C^{N N}_{T,X},\;X=L,R$~\cite{Cirigliano:2017azj,Davidson:2017nrp}.
The overlap integrals,
\begin{align}
D &= \frac{4}{\sqrt{2}} m_\mu \int_0^\infty \dd r \, r^2 \left[ - E(r)\right] \left( g_e^- f_\mu^- + f_e^- g_\mu^-\right) ,\\
S^{(p)} &= \frac{1}{2\sqrt{2}}  \int_0^\infty \dd r \, r^2 Z \rho^{(p)} \left( g_e^- g_\mu^- - f_e^- f_\mu^-\right) ,\\
S^{(n)} &= \frac{1}{2\sqrt{2}}  \int_0^\infty \dd r \, r^2 N \rho^{(n)} \left( g_e^- g_\mu^- - f_e^- f_\mu^-\right) ,\\
V^{(p)} &= \frac{1}{2\sqrt{2}}  \int_0^\infty \dd r \, r^2 Z \rho^{(p)} \left( g_e^- g_\mu^- + f_e^- f_\mu^-\right) ,\\
V^{(n)} &= \frac{1}{2\sqrt{2}}  \int_0^\infty \dd r \, r^2 N \rho^{(n)} \left( g_e^- g_\mu^- + f_e^- f_\mu^-\right) ,
\end{align}
contain all the information about the structure of the nucleus, here assumed to be spherically symmetric, through the density distributions of charge $\rho^{(c)}$, protons $\rho^{(p)}$, and neutrons $\rho^{(n)}$. These are normalized via
\begin{align}
 \int_0^\infty \dd r\, 4\pi r^2 \rho^{(c),(p),(n)}(r)=1 \,.
\end{align}
 The radial electric field $E(r)$, relevant for the dipole overlap integral $D$, is defined as
\begin{align}
E(r) = \frac{Z e}{r^2} \int_0^r\dd \tilde{r}\, \tilde{r}^2 \rho^{(c)}(\tilde{r}) \,.
\end{align}
The functions $g_e^-$, $f_e^-$, $g_\mu^-$, and $f_\mu^-$ are radial parts of the electron and muon relativistic wavefunctions, determined by numerically solving the relevant Dirac equations in the external electric field of the nucleus, following Ref.~\cite{Heeck:2021adh}.

The overlap integrals can be calculated for a given isotope after specifying the nuclear distributions. The Wilson coefficients entering Eq.~\eqref{eq:BR} can be obtained in a given new-physics model through standard procedure of matching at the new-physics scale and running down the coefficients to the experimental scale via renormalization-group equations. Here, we will take them to be arbitrary input parameters.

\section{Nuclear distributions}

It is difficult to compute nuclear charge distributions accurately from the first principles \cite{Lapoux:2016exf,Arthuis:2020toz,Malbrunot-Ettenauer:2021fnr}, especially for heavy elements. Instead, they can be extracted from experiments. Information about the nuclear distributions $\rho$ can be obtained via spectroscopy in (muonic) atoms and through elastic scattering. Relying on electromagnetic interactions, this gives access to the charge distribution $\rho^{(c)}$, for which numerous data tables exist~\cite{DeVries:1987atn,Fricke:1995zz}. Spectroscopic measurements typically allow only to extract the value of the  root-mean-square charge radius. Electron--nucleus scattering data can probe electromagnetic form-factors at different values of the momentum transfer, but available data points are sparse.  

A model-independent determination of the charge  distribution is practically impossible and is typically replaced by fitting a theoretically or phenomenologically motivated ansatz to data.
Widely adopted parametrizations for spherically symmetric charge distributions with varying degrees of complexity are listed below:
\begin{enumerate}
\item Three-parameter Fermi model (3pF)~\cite{Yennie:1954zz,Hahn:1956zz}:
\begin{align}
\rho^{(c)} (r) = \frac{\rho_0 }{1 + \exp\left(\frac{r-c}{z}\right)}\left(1 + w\frac{ r^2}{c^2}\right) .
\end{align}
The two-parameter Fermi model (2pF) can be obtained as the special case $w=0$; the one-parameter Fermi function (1pF) is defined here through $w=0$ and $z = \unit[0.52]{fm}$ (which corresponds to a constant surface thickness of $\unit[2.3]{fm}$). 
\item Three-parameter Gaussian model (3pG)~\cite{Hahn:1956zz}:
\begin{align}
\rho^{(c)} (r) = \frac{\rho_0 }{1 + \exp\left(\frac{r^2-c^2}{z^2}\right)}\left(1 + w\frac{ r^2}{c^2}\right) .
\end{align}
\item Modified-harmonic oscillator model (MHO)~\cite{Hofstadter:1957wk}:
\begin{align}
\rho^{(c)} (r) = \rho_0 \left(1 + w\frac{ r^2}{a^2}\right) \text{e}^{-r^2/a^2} .
\end{align}
\item Fourier--Bessel expansion (FB)~\cite{Dreher:1974pqw}:
\begin{align}
\rho^{(c)} (r) = \begin{cases}
\sum_k^n a_k j_0 \left(\frac{k \pi r}{R}\right) , & \, \, r\leq R\,,\\
0\,, & \,\, r>R\,.
\end{cases}
\end{align}
Here, $a_{1,\dots,n}$ are the FB coefficients and $j_0(x) = \sin(x)/x$ the spherical Bessel function of order zero.
$n\leq 18$ for our data.
\item Sum of Gaussians (SOG)~\cite{Sick:1974suq}:
\begin{align}
\rho^{(c)} (r) = 
\sum_k^n Q_k\,\frac{ \text{e}^{-\frac{(r-R_k)^2}{\gamma^2}}+\text{e}^{-\frac{(r+R_k)^2}{\gamma^2}}}{2\pi^{3/2} \gamma^3 (1+ 2 R_k^2/\gamma^2)}\,,
\end{align}
with $\sum_k^n Q_k = 1$.
\end{enumerate}

The rudimentary 1pF parametrization depends only on a single parameter: the nuclear charge radius. This parameter is given for all elements and isotopes that we consider here in Ref.~\cite{Angeli:2013epw}. In many cases, this is the only available parametrization, making it extremely valuable despite its lack of substructure.
For the multi-parameter parametrizations we use the available data tables from Refs.~\cite{DeVries:1987atn,Boeglin:1988yfc,Fricke:1995zz,Wesseling:1997zz,Kabir:2015igz}, generally choosing the newest possible data set for a given isotope.

Not every source provides uncertainties for the charge distribution's parameters, and in some instances the uncertainties are possibly underestimated. For example, the typical errors on the charge radius relevant for 1pF are small, which does however not imply accurate knowledge of the charge distribution. Moreover, muon-to-electron conversion is sensitive to short distance structure of the distributions; thus, to obtain  reliable estimates  for the uncertainty on the overlap integrals, we perform computations for several available parametrizations for a given isotope and then compare the results. 

We restrict our survey to the 236 stable isotopes with natural abundance above $1\%$. Only half of these isotopes have a charge parametrization with more than two parameters, for the other half we have to rely on the simplistic 1pF model.

The charge distribution is the relevant quantity to calculate the muon and electron wavefunctions~\cite{Heeck:2021adh} as well as the dipole overlap integral $D$.
For the overlap integrals $S^{(p)}$ and $V^{(p)}$, the distribution of \emph{protons} is needed. This distribution is not identical with the charge distribution due to the finite proton size.
Proton distributions are available in Ref.~\cite{Patterson:2003wbr} for 3pF, MHO, and 3pG parametrizations, not for the more-realistic SOG and FB forms. Of course, to leading order we have $\rho^{(p)} = \rho^{(c)}$, the approximation used in the past when calculating overlap integrals.
Translating $\rho^{(c)}$ to $\rho^{(p)}$ by folding-in the proton's charge distribution is at best an approximate solution but provides a useful estimate for the size of this effect. We follow Ref.~\cite{Patterson:2003wbr}  [specifically their Eqs.~(6) and (7)] to translate our 1pF charge distribution into a proton distribution $\rho^{(p)}$.

Finally, for $S^{(n)}$ and $V^{(n)}$ we require the distribution of \emph{neutrons} in the nucleus, which is considerably more difficult to ascertain experimentally. Pion scatting has been used in some cases to obtain the distributions of protons and neutrons, from which we can derive $\rho^{(n)}$.
A decent approximation used in the past is simply  $\rho^{(n)} = \rho^{(p)}$, which is however expected to break down in heavy nuclei with $N \gg Z$, where the neutron skin extends further out.
If $\mu\to e$ conversion should be observed, it would be highly desirable to experimentally improve the nuclear distributions in the relevant nuclei in order to reduce the uncertainty and allow for a determination of the relevant Wilson coefficients in Eq.~\eqref{eq:BR}.

\section{Overlap integrals}

For a given spherical charge distribution $\rho^{(c)}$ we compute the electric potential by the well-known expression
\begin{equation}
    V(r) = - 4\pi Z \alpha\left[\int_0^r \dd z \frac{z^2}{r}\rho^{(c)}(z)+ \int_r^\infty \dd z \,z\rho^{(c)}(z)\right]
\end{equation}
and then use it to solve the Dirac equations numerically using a fourth-order Runge--Kutta method to extract the muon and electron wave functions~\cite{Heeck:2021adh}, which are then inputted into the overlap integrals.

\subsection{Dipole integral \texorpdfstring{$D$}{D}}

\begin{figure}
\includegraphics[width=0.48\textwidth]{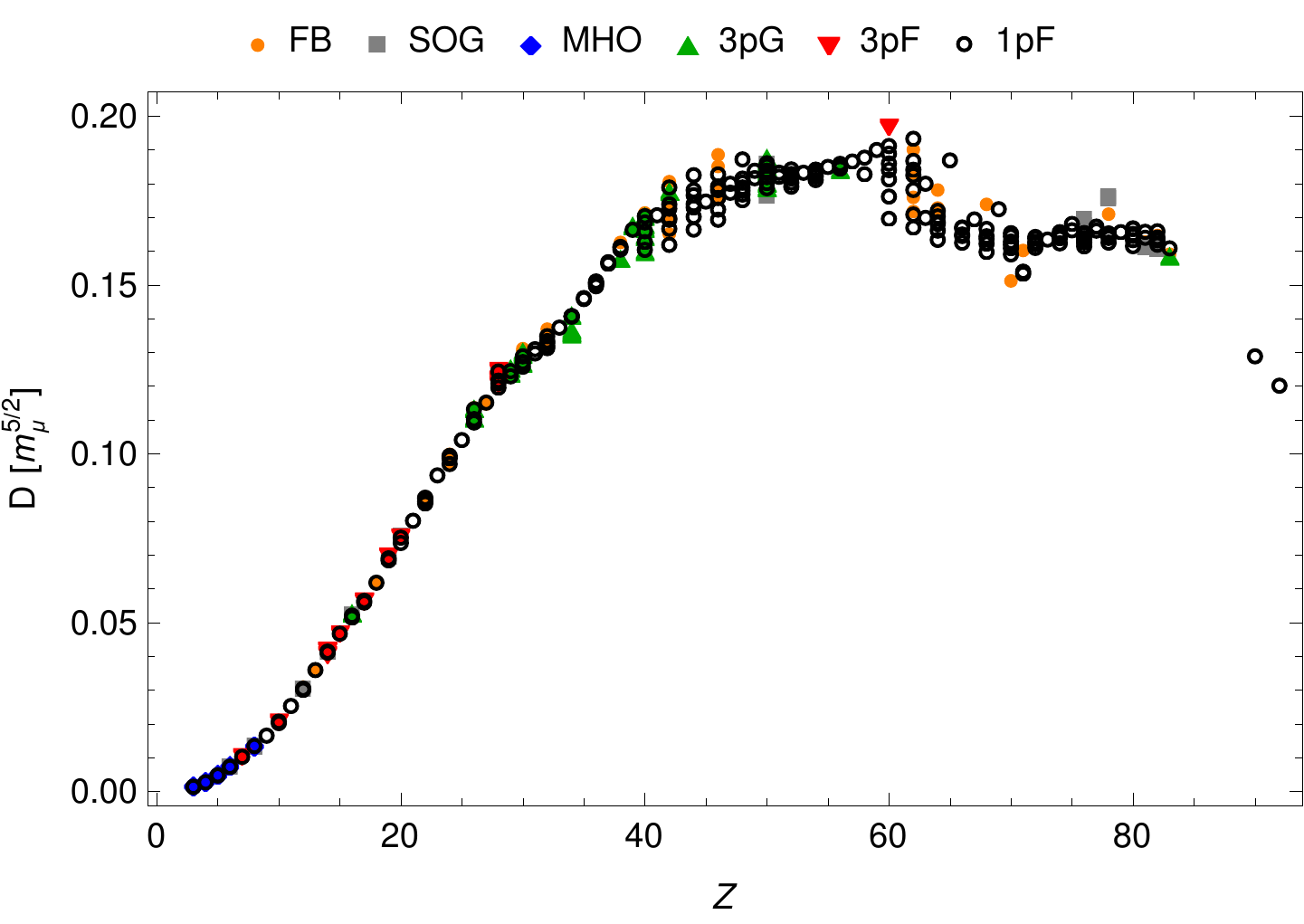}\vspace{4ex}
\includegraphics[width=0.48\textwidth]{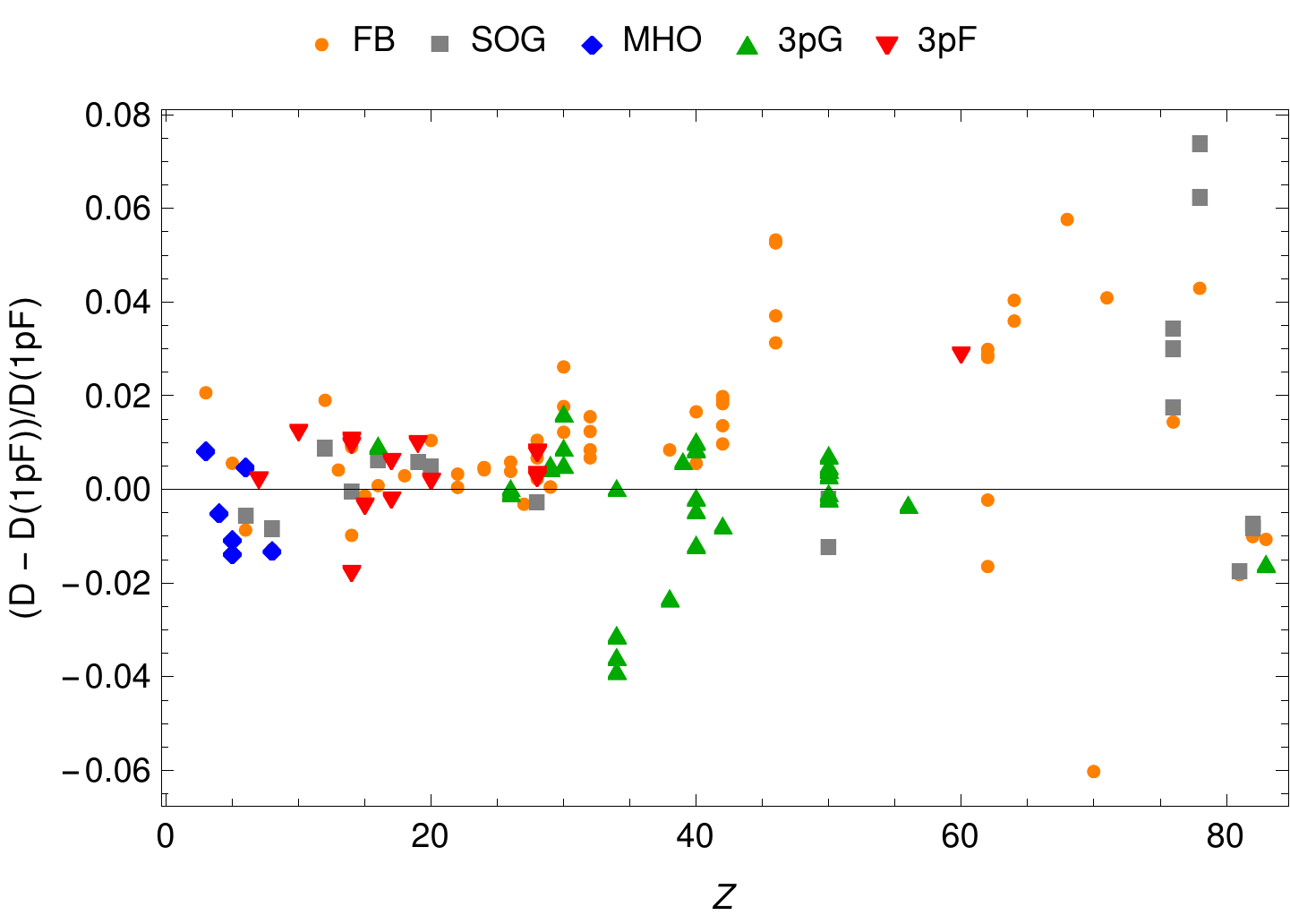}\vspace{4ex}
\includegraphics[width=0.48\textwidth]{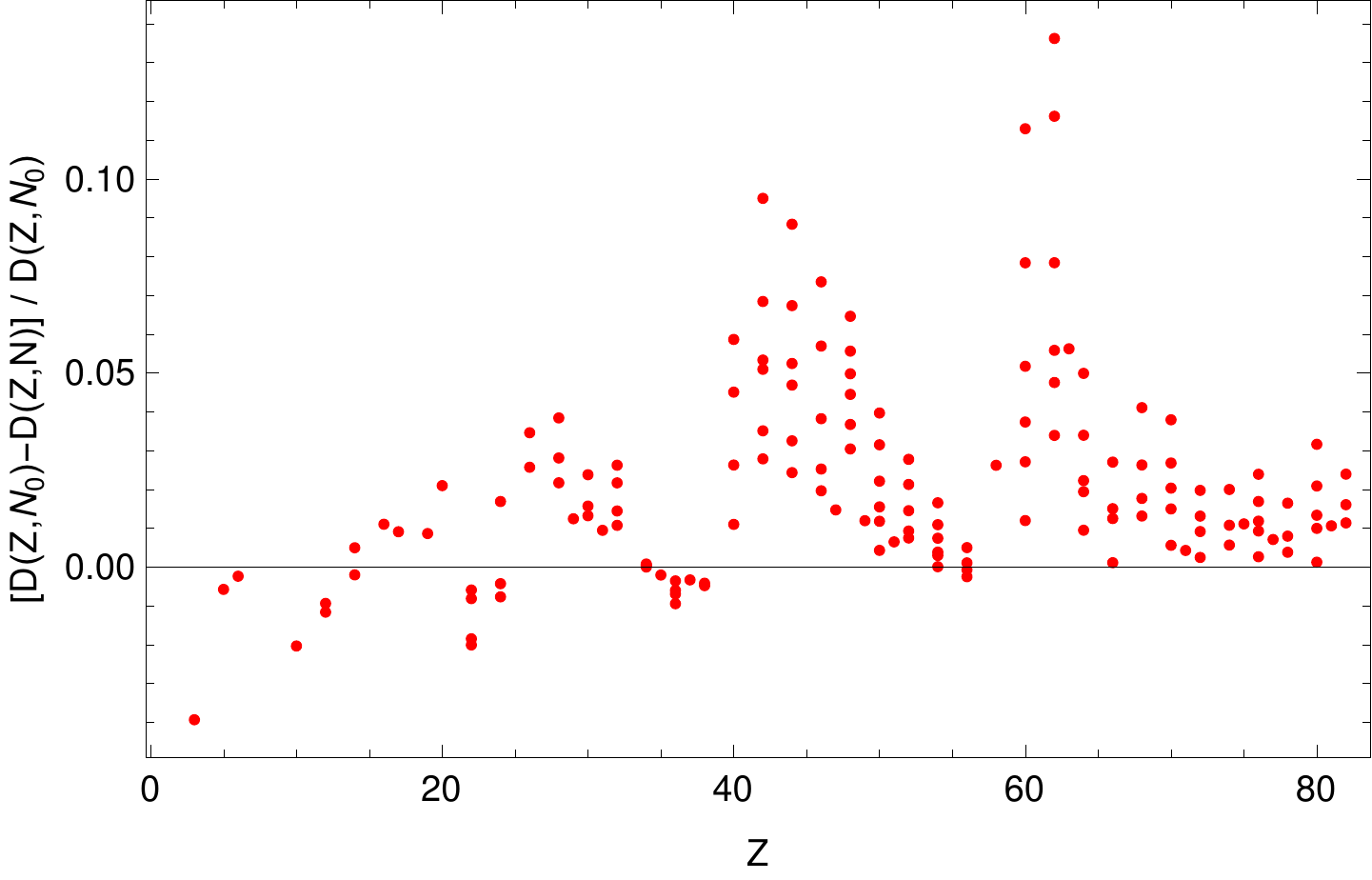}
\caption{
Top: dipole overlap integral $D$ vs $Z$ for all available parametrizations.
Middle: difference of $D$ calculated with the 1pF charge distribution and other distributions.
Bottom: isotope dependence calculated using 1pF distributions, i.e.~the difference in $D$ between isotopes $(Z,N)$ and $(Z,N_0)$, with $N_0$ the smallest stable neutron number.
\label{fig:D_coefficient}}
\end{figure}

Our results for the dipole overlap integral $D$ as a function of $Z$ are shown in Fig.~\ref{fig:D_coefficient} (top) for all elements, isotopes, and available parametrizations. The values are also listed in Tab.~\ref{tab:results}.
$D$ is the overlap integral with the smallest uncertainties since it depends exclusively on the charge distribution of the isotope, which is experimentally accessible through electron scattering and spectroscopy. In the small $Z$ region, $Z<30$, the uncertainty due to the charge distribution is less than $2\%$, but grows to $8\%$ for large $Z$. 
The isotope dependence easily exceeds this uncertainty at medium $Z$ and should not be neglected.

\subsection{Overlap integral \texorpdfstring{$S^{(p)}$}{Sp}}

\begin{figure}[th]
\includegraphics[width=0.48\textwidth]{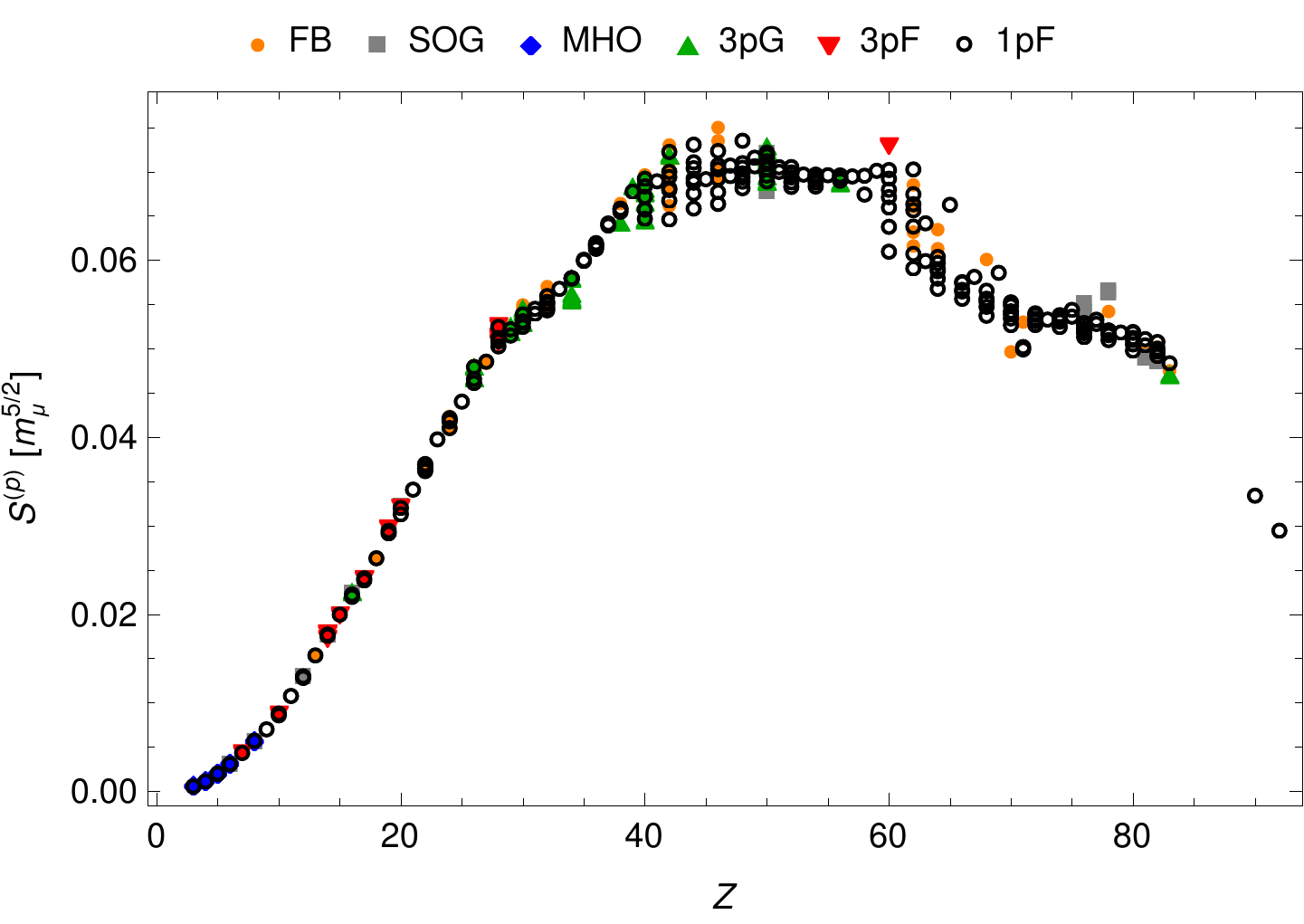}\vspace{4ex}
\includegraphics[width=0.48\textwidth]{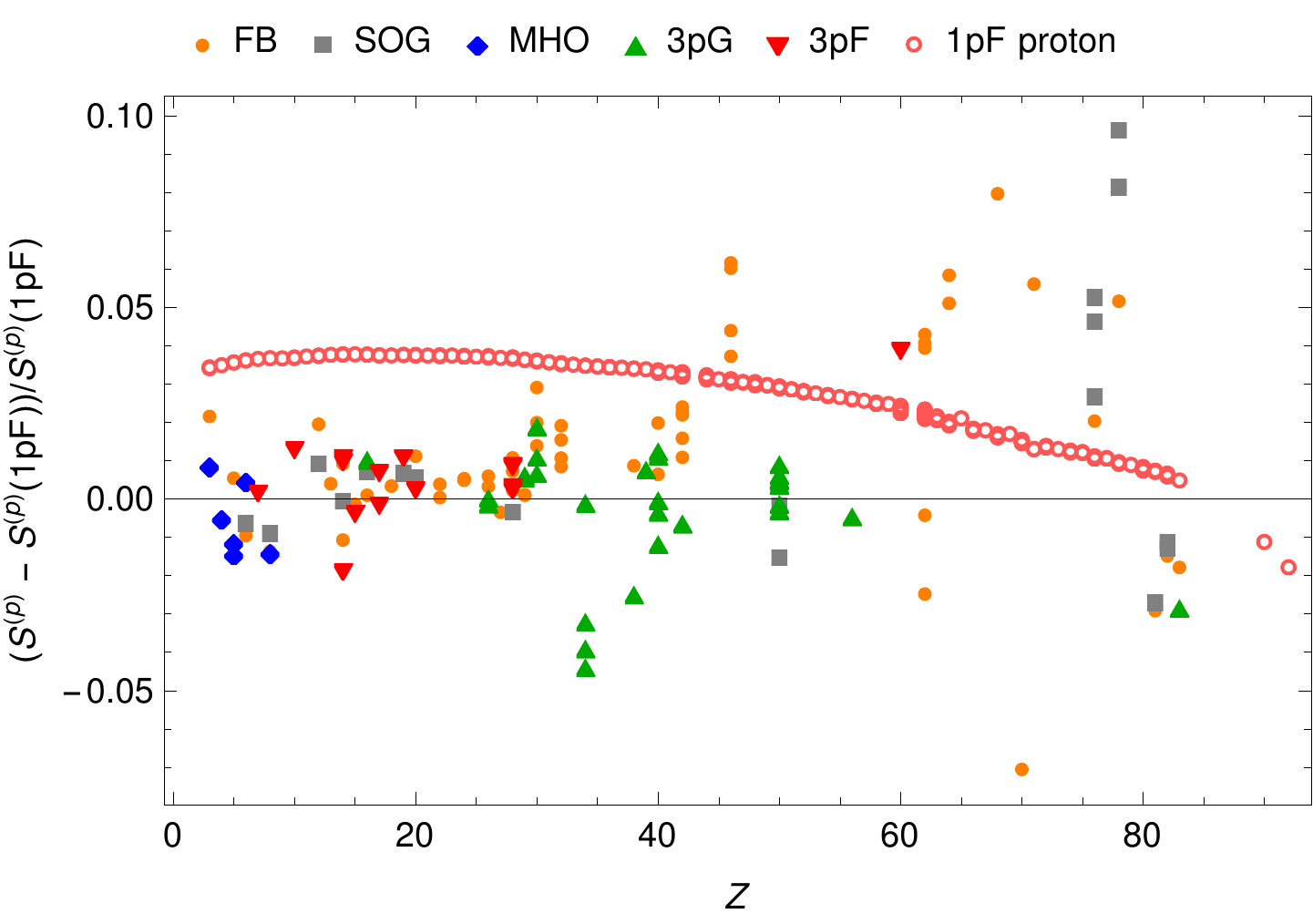}\vspace{4ex}
\includegraphics[width=0.48\textwidth]{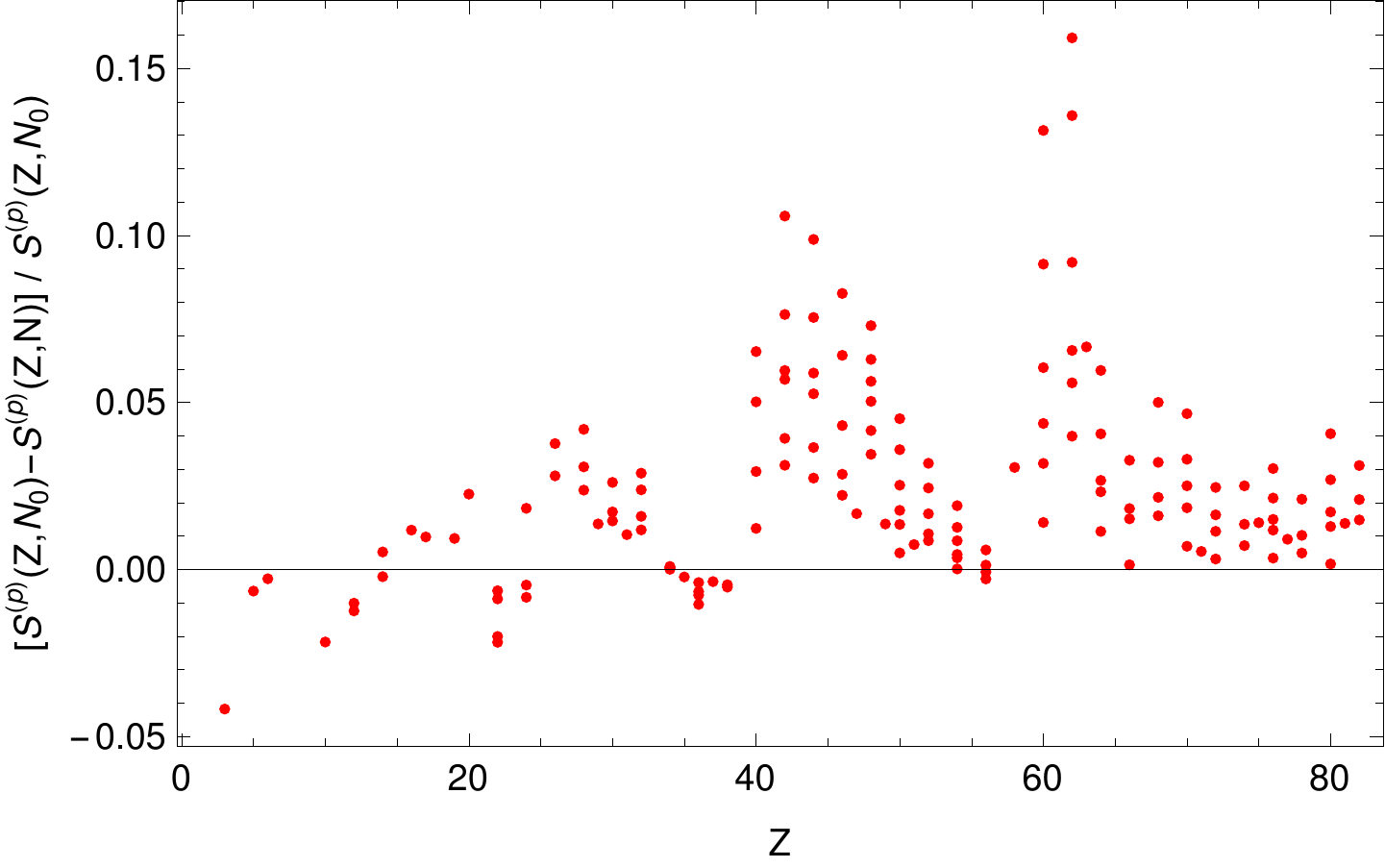}\vspace{4ex}
\caption{
Same as Fig.~\ref{fig:D_coefficient} but for the overlap integral $S^{(p)}$.
In the middle plot, 1pF proton refers to the proton-center distribution that is obtained by unfolding the proton size from the 1pF charge distribution, following Ref.~\cite{Patterson:2003wbr}.
\label{fig:Sp_coefficient}}
\end{figure}

The results for the overlap integral $S^{(p)}$ are shown in Fig.~\ref{fig:Sp_coefficient}.
Assuming $\rho^{(p)} = \rho^{(c)}$, the uncertainty in the low-$Z$ regime due to the charge-distribution uncertainty is around $2\%$; using instead the proton distribution $\rho^{(p)}$ that approximately accounts for the proton size systematically increases $S^{(p)}$ by around $3.8\%$, independent of the number of neutrons.
We take these values to be more realistic for $S^{(p)}$ and estimate an uncertainty of $5\%$.

For larger $Z$, the difference between charge and proton distribution becomes negligible and the uncertainty is dominated by the charge-distribution parametrization, which reaches up to $10\%$.
 The isotope dependence at large $Z$ is non-negligible, reaching up to $16\%$ for samarium, and does not depend on whether we use $\rho^{(p)} = \rho^{(c)}$ or an unfolded $\rho^{(p)}$.

To provide a comparison of our rather simplistic $\rho^{(p)}$, we use the \Element{Ti}{22}{50} results from Ref.~\cite{Yang:2019pbx} for $\rho^{(p)}$, obtained via state-of-the-art mean-field models calibrated to data; this yields the overlap integral $S^{(p)}/m_\mu^{5/2} = 0.0392$, compared to 1pF results of $S^{(p)}/m_\mu^{5/2} = 0.0384$. This 2\% deviation is well within our estimated error bars and illustrates the accuracy of our results.
For \Element{Pb}{82}{208}, we similarly use the more realistic state-of-the-art calculation of $\rho^{(p)}$ from Ref.~\cite{Kim:2021skf}, which gives $S^{(p)}/m_\mu^{5/2} = 0.0476$, compared to $S^{(p)}/m_\mu^{5/2} = 0.0495$ obtained using our 1pF parametrization. The 4\% deviation is again well within our error bars.

\subsection{Overlap integral \texorpdfstring{$V^{(p)}$}{Vp}}

The results for the overlap integral $V^{(p)}$ are qualitatively and even quantitatively so similar to  $S^{(p)}$ that we omit a figure analogous to Fig.~\ref{fig:Sp_coefficient}.
For small $Z$, we have $V^{(p)}\simeq S^{(p)}$~\cite{Kitano:2002mt}.
The uncertainties are the same as for $S^{(p)}$: $5\%$ for $Z<30$ and $10\%$ for $Z\geq 30$, and we again take the results of the unfolded proton distribution as a more realistic estimate that improves upon the approximation $\rho^{(p)} = \rho^{(c)}$.

We again compare our results with more sophisticated theoretical $\rho^{(p)}$ models for \Element{Ti}{22}{50}~\cite{Yang:2019pbx} and \Element{Pb}{82}{208}~\cite{Kim:2021skf}, which only differ from our Tab.~\ref{tab:results} values by 2\% and 3\%, respectively.

\subsection{Overlap integrals \texorpdfstring{$S^{(n)}$}{Sn} and \texorpdfstring{$V^{(n)}$}{Vn}}

The overlap integrals $S^{(n)}$ and $V^{(n)}$ depend on the neutron density $\rho^{(n)}$, which is far more difficult to measure than the proton or charge density and only known for very few nuclei.
We have no choice but to make a theoretical ansatz here.
The elementary approximation is simply $\rho^{(n)}\simeq \rho^{(p)}$, in which case $S^{(n)} \simeq (N/Z) S^{(p)}$ and $V^{(n)} \simeq (N/Z) V^{(p)}$, listed in Tab.~\ref{tab:results}.

For comparison to more realistic distributions, we use the state-of-the-art predictions of $\rho^{(n)}$ for \Element{Ti}{22}{50}~\cite{Yang:2019pbx} and \Element{Pb}{82}{208}~\cite{Kim:2021skf} as well as the experimental data for \Element{Ca}{20}{40}~\cite{Zenihiro:2018rmz} to calculate $S^{(n)}$ and $V^{(n)}$. In all cases, the deviations are between 2\% and 3\%, well within our allotted errors. Despite the many approximations we had to make to calculate the neutron overlap integrals, the results seem robust.

\subsection{Summary}

\begin{figure}[th]
\includegraphics[width=0.48\textwidth]{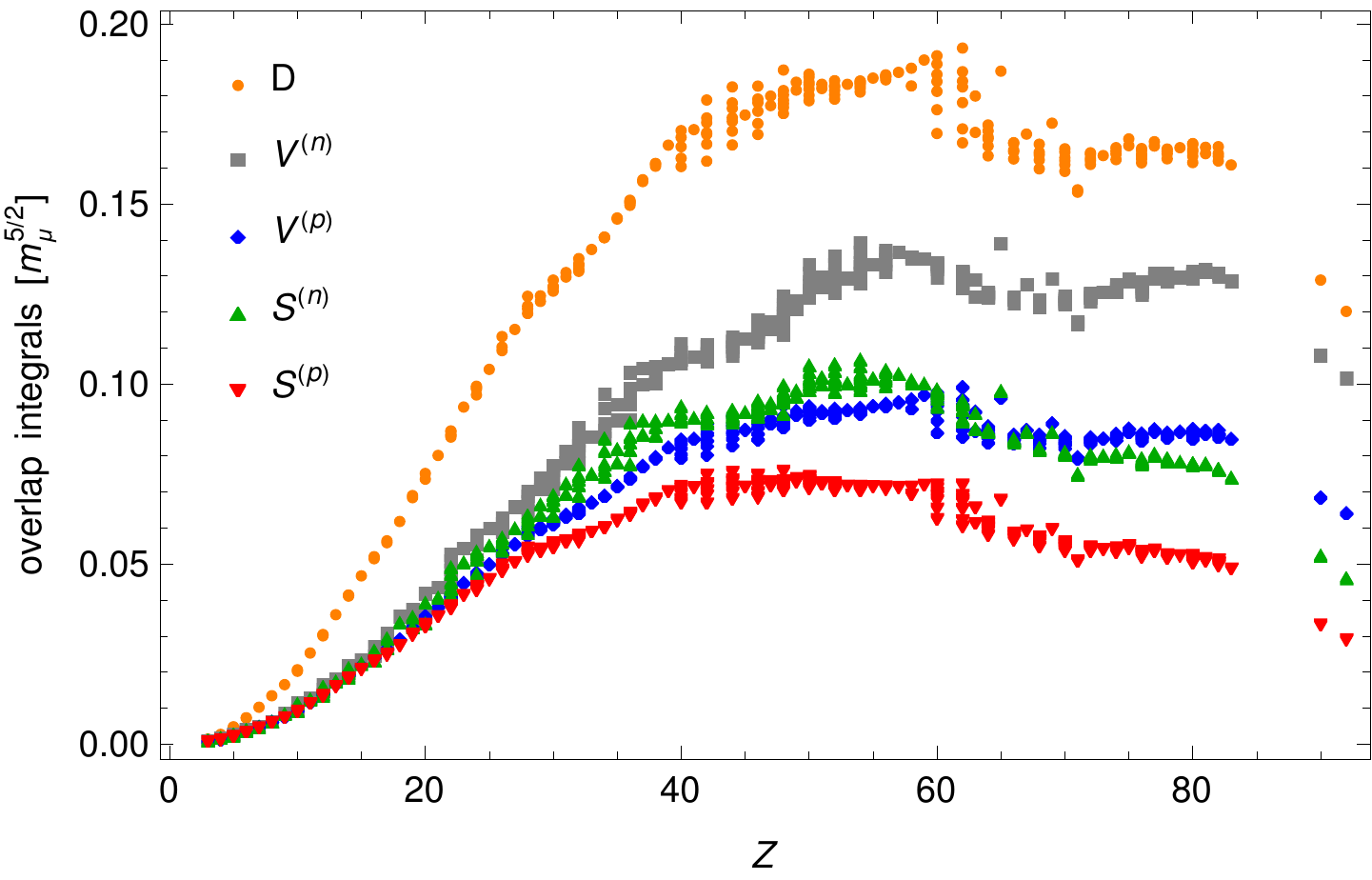}
\caption{
All overlap integrals as a function of $Z$, using the data from Tab.~\ref{tab:results}. We approximate $\rho^{(n)} = \rho^{(p)}$ and use the proton-center distribution unfolded from the 1pF charge distribution.
\label{fig:all_overlaps}}
\end{figure}

Our final results are shown in Fig.~\ref{fig:all_overlaps} and tabulated in Tab.~\ref{tab:results}.
For $D$, they are based on the 1pF charge distribution and we estimate an uncertainty of about $2\%$ at low $Z$ that grows to $8\%$ at large $Z$.
For $S^{(p)}$ and $V^{(p)}$, we unfold the 1pF charge distribution to account for the finite proton size and obtain the proton-center distribution following Ref.~\cite{Patterson:2003wbr}. We estimate the uncertainties to range from $5\%$ for low $Z$ to $10\%$ at high $Z$.
For the neutron overlap integrals $S^{(n)}$ and $V^{(n)}$ we approximate $\rho^{(n)} = \rho^{(p)}$ and use the same distributions as for $S^{(p)}$ and $V^{(p)}$. For small $Z$, we expect a similar uncertainty around $5\%$, while at large $Z$ the approximation $\rho^{(n)} = \rho^{(p)}$ is going to become increasingly worse, introducing an error in excess of $10\%$.
While the accuracy of the overlap integrals is sufficient for our current purposes, efforts should be undertaken to improve them if muon-to-electron conversion is observed.

\section{Complementarity of targets}

Assuming optimistically that COMET or Mu2e observe $\mu\to e$ conversion on their aluminium target with sufficient significance to claim a discovery, what target nucleus should be investigated next?
One answer was given long ago: any high-$Z$ target, e.g.~gold, since the overlap integrals become more distinguishable at large $Z$~\cite{Kitano:2002mt,Cirigliano:2009bz,Bartolotta:2017mff,Davidson:2017nrp,Davidson:2018kud,Davidson:2020hkf} (see Fig.~\ref{fig:all_overlaps}). Unfortunately, high-$Z$ elements have a very short muon lifetime that renders them difficult to use in experimental setups similar to Mu2e. 

In order to find appropriate \emph{low}-$Z$ targets, we must first find a quantitative measure of complementarity.
Following Refs.~\cite{Davidson:2017nrp,Davidson:2018kud}, we write the spin-independent branching ratio from Eq.~\eqref{eq:BR} as
\begin{align}
\BR_\text{SI}= \frac{32 G_F^2  }{\Gamma_\text{capture}} \left[  
|\vec{v}\cdot \vec{C}_L|^2+|\vec{v}\cdot \vec{C}_R|^2\right] ,
\label{eq:BR2}
\end{align}
where
\begin{align}
\vec{v} \equiv \left( \frac{D}{4}, V^{(p)}, S^{(p)}, V^{(n)}, S^{(n)}\right) 
\label{eq:v}
\end{align}
is a vector specific to the $\mu\to e $ conversion target and 
\begin{align}
\vec{C}_{L} \equiv \left(C_{D,R}, C^{pp}_{V,L}, C^{pp}_{S,R}, C^{nn}_{V,L}, C^{nn}_{S,R}\right) , 
\label{eq:C}
\end{align}
(similar for $\vec{C}_{R} $) contains all new-physics information. By measuring $\mu\to e$ conversion on different nuclei we can measure $\vec{C}$ along different directions in order to determine its individual components. Since all overlap integrals are positive and of similar magnitude, the different vectors $\vec{v}$ all point roughly in the same direction, so this procedure requires a precise understanding of the nuclear structure.

\begin{figure}
\includegraphics[width=0.48\textwidth]{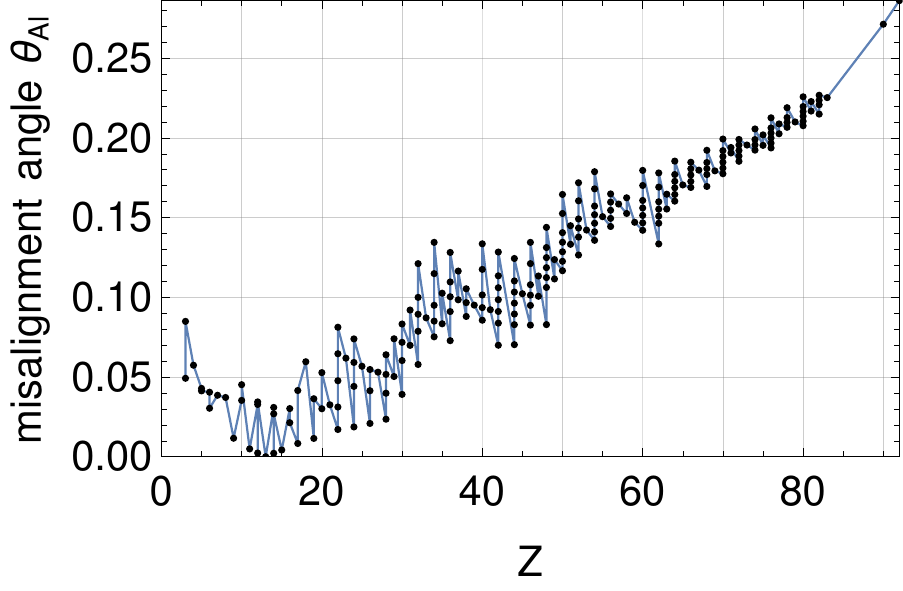}
\includegraphics[width=0.48\textwidth]{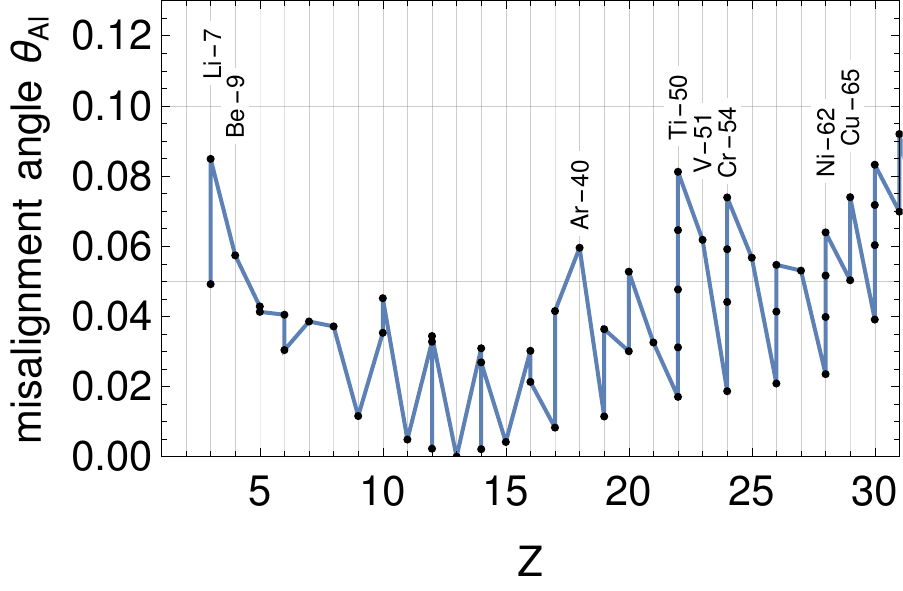}
\caption{
Misalignment angle with Al, as calculated with Eq.~\eqref{eq:theta} using our data from Tab.~\ref{tab:results}. The misalignment angle increases with the number of neutrons in isotopes.
\label{fig:Al_misalignment}}
\end{figure}

Assuming $\mu\to e $ conversion is measured on Al, the next target material should be chosen so as to provide as much complementary information to Al as possible, which is equivalent to demanding that the corresponding vector $\vec{v}$ is maximally misaligned with $\vec{v}_\text{Al}$. This can be quantified through the misalignment angle\footnote{Since the dipole overlap integral $D$ is somewhat special and in any case well constrained through $\mu\to e \gamma$, one might consider defining $\vec{v}$ without $D$; this turns out to not make a difference in $\theta$, which is dominated by the proton vs.~neutron difference and essentially insensitive to the $D$ direction.}
\begin{align}
\theta_\text{Al} = \arccos \left( \frac{\vec{v}\cdot \vec{v}_\text{Al}}{|\vec{v}|| \vec{v}_\text{Al}|}\right) .
\label{eq:theta}
\end{align}
We show this angle in Fig.~\ref{fig:Al_misalignment}, which clearly confirms that high-$Z$ targets have the largest complementarity with Al overall. 

Restricting ourselves to Mu2e-friendly $Z < 25$ targets, lithium-7 and titanium-50 show the largest complementarity with respect to aluminium, followed by chromium-54 and vanadium. They have larger $N/Z $ ratios, $ 2.33$ and $2.27$ for lithium-7 and titanium-50, respectively, compared to Al's $N/Z\simeq 2.08$, which might ultimately help to distinguish CLFV operators involving protons from those involving neutrons~\cite{Davidson:2018kud}. 
Lithium has already been identified as a promising target in Ref.~\cite{Davidson:2018kud}.
Titanium has long been proposed as a suitable second target for aluminum-based experiments, and our analysis shows that the isotope Ti-50 would be particularly useful; aside from the conversion rate and the background from muon decay in orbit, different isotopes of an element are expected to behave essentially identically experimentally, notably because the conversion energy depends only weakly on the number of neutrons~\cite{Heeck:2021adh}.
The theoretically interesting isotopes Ti-50, Ti-49, and Cr-54 have a low natural abundance and are difficult enrich in the large quantities necessary for conversion experiments; Li-7 and V-51, on the other hand, are the dominant isotopes and hence practically preferable as second targets after an observation of $\mu\to e$ conversion on aluminium.

\section{Conclusions}

The search for lepton flavor violation is one of our most sensitive probes of physics beyond the Standard Model. Experiments searching for $\mu^-\to e^-$ conversion such as COMET, DeeMe, and Mu2e, promise to improve existing limits by several orders of magnitude. Robust theoretical predictions, as presented here, are crucial ingredients for experimental simulations of possible signal strength and are relevant for the choice of alternative targets.   An observation of the coherent conversion signal would clearly indicate new physics. Still, it would not provide enough information to understand the nature of the new interactions. Our results allow tracking the nucleus-dependence of the $\mu^-\to e^-$ conversion rate by looking at different target materials, which would then help to discriminate the possible underlying new-physics models and effective operators. Such studies are instrumental in the context of proposed upgrades of the already approved experiments~\cite{Mu2e:2018osu}. 

Our results indicate that the isotope dependence can exceed the uncertainty due to the nuclear charge distribution. Thus, experiments must carefully control the isotope composition of the targets to enable the proper interpretation of the results in terms of bounds on underlying short distance parameters of the effective Lagrangians. The isotope dependence can also improve the experiments' potential to distinguish various New Physics scenarios if a signal is observed. 

Further improvement of the total coherent conversion rates requires more precise determination of the proton, neutron, and charge density profiles. The progress in the many-body computational methods may allow in the future ab-initio evaluation of these density functions, which would be highly desirable for experimentally studied target materials.

\section*{Acknowledgements}
We thank all members of the Mu2e-II Snowmass21 Group for valuable discussions and support, especially Frank Porter, Lorenzo Calibbi, David Hitlin, Sophie Middleton, and Leo Borrell.
This work was partly supported by JSPS KAKENHI Grant Numbers JP18H01210 and JP21H00081 (Y.U.). R.S.~is supported by the United States Department of Energy under Grant Contract DE-SC0012704.

\appendix

\clearpage

\onecolumngrid

\newgeometry{top=2cm,bottom=5cm}
\section{Table of our results}
\label{app:table}

\centering

\bottomcaption{Table of our results, showing for each stable isotope with $Z\geq 3$ and natural abundance above $1\%$ the overlap integrals in units of $m_\mu^{5/2}$.\label{tab:results}}
\tablehead{   & $D [m_\mu^{5/2}]$ & $S^{(p)} [m_\mu^{5/2}]$  & $V^{(p)} [m_\mu^{5/2}]$ & $S^{(n)} [m_\mu^{5/2}]$ & $V^{(n)} [m_\mu^{5/2}]$ \\
 \hline}
\begin{supertabular}{llllll}
\Element{Li}{3}{6} & $0.0013$ & $0.00056$ & $0.000564$ & $0.00056$ & $0.000564$ \\
\Element{Li}{3}{7} & $0.00135$ & $0.000584$ & $0.000588$ & $0.000779$ & $0.000783$ \\
\Element{Be}{4}{9} & $0.00268$ & $0.00116$ & $0.00117$ & $0.00145$ & $0.00146$ \\
\Element{B}{5}{10} & $0.00473$ & $0.00206$ & $0.00208$ & $0.00206$ & $0.00208$ \\
\Element{B}{5}{11} & $0.00475$ & $0.00208$ & $0.0021$ & $0.00249$ & $0.00251$ \\
\Element{C}{6}{12} & $0.00725$ & $0.00318$ & $0.00322$ & $0.00318$ & $0.00322$ \\
\Element{C}{6}{13} & $0.00727$ & $0.00319$ & $0.00323$ & $0.00372$ & $0.00376$ \\
\Element{N}{7}{14} & $0.0102$ & $0.00449$ & $0.00456$ & $0.00449$ & $0.00456$ \\
\Element{O}{8}{16} & $0.0134$ & $0.00591$ & $0.00601$ & $0.00591$ & $0.00601$ \\
\Element{F}{9}{19} & $0.0165$ & $0.00726$ & $0.00743$ & $0.00807$ & $0.00825$ \\
\Element{Ne}{10}{20} & $0.0202$ & $0.00891$ & $0.00914$ & $0.00891$ & $0.00914$ \\
\Element{Ne}{10}{22} & $0.0206$ & $0.0091$ & $0.00934$ & $0.0109$ & $0.0112$ \\
\Element{Na}{11}{23} & $0.0253$ & $0.0112$ & $0.0115$ & $0.0122$ & $0.0125$ \\
\Element{Mg}{12}{24} & $0.03$ & $0.0133$ & $0.0137$ & $0.0133$ & $0.0137$ \\
\Element{Mg}{12}{25} & $0.0304$ & $0.0135$ & $0.0139$ & $0.0146$ & $0.0151$ \\
\Element{Mg}{12}{26} & $0.0303$ & $0.0134$ & $0.0139$ & $0.0157$ & $0.0162$ \\
\Element{Al}{13}{27} & $0.0359$ & $0.0159$ & $0.0165$ & $0.0172$ & $0.0178$ \\
\Element{Si}{14}{28} & $0.0413$ & $0.0183$ & $0.019$ & $0.0183$ & $0.019$ \\
\Element{Si}{14}{29} & $0.0414$ & $0.0184$ & $0.0191$ & $0.0197$ & $0.0204$ \\
\Element{Si}{14}{30} & $0.0411$ & $0.0182$ & $0.0189$ & $0.0208$ & $0.0217$ \\
\Element{P}{15}{31} & $0.0467$ & $0.0207$ & $0.0216$ & $0.0221$ & $0.0231$ \\
\Element{S}{16}{32} & $0.052$ & $0.0231$ & $0.0242$ & $0.0231$ & $0.0242$ \\
\Element{S}{16}{34} & $0.0515$ & $0.0228$ & $0.0239$ & $0.0257$ & $0.0269$ \\
\Element{Cl}{17}{35} & $0.0564$ & $0.025$ & $0.0263$ & $0.0264$ & $0.0278$ \\
\Element{Cl}{17}{37} & $0.0559$ & $0.0247$ & $0.026$ & $0.0291$ & $0.0306$ \\
\Element{Ar}{18}{40} & $0.0618$ & $0.0273$ & $0.0289$ & $0.0334$ & $0.0353$ \\
\Element{K}{19}{39} & $0.069$ & $0.0305$ & $0.0324$ & $0.0321$ & $0.0341$ \\
\Element{K}{19}{41} & $0.0684$ & $0.0302$ & $0.0321$ & $0.035$ & $0.0372$ \\
\Element{Ca}{20}{40} & $0.0751$ & $0.0332$ & $0.0354$ & $0.0332$ & $0.0354$ \\
\Element{Ca}{20}{44} & $0.0736$ & $0.0325$ & $0.0346$ & $0.039$ & $0.0416$ \\
\Element{Sc}{21}{45} & $0.0801$ & $0.0354$ & $0.0379$ & $0.0404$ & $0.0433$ \\
\Element{Ti}{22}{46} & $0.0852$ & $0.0375$ & $0.0404$ & $0.0409$ & $0.0441$ \\
\Element{Ti}{22}{47} & $0.0857$ & $0.0378$ & $0.0406$ & $0.0429$ & $0.0462$ \\
\Element{Ti}{22}{48} & $0.0859$ & $0.0379$ & $0.0407$ & $0.0448$ & $0.0481$ \\
\Element{Ti}{22}{49} & $0.0868$ & $0.0383$ & $0.0412$ & $0.047$ & $0.0505$ \\
\Element{Ti}{22}{50} & $0.0869$ & $0.0384$ & $0.0412$ & $0.0488$ & $0.0525$ \\
\Element{V}{23}{51} & $0.0936$ & $0.0413$ & $0.0445$ & $0.0502$ & $0.0542$ \\
\Element{Cr}{24}{50} & $0.0986$ & $0.0434$ & $0.047$ & $0.047$ & $0.0509$ \\
\Element{Cr}{24}{52} & $0.0994$ & $0.0437$ & $0.0474$ & $0.051$ & $0.0553$ \\
\Element{Cr}{24}{53} & $0.099$ & $0.0436$ & $0.0472$ & $0.0526$ & $0.0571$ \\
\Element{Cr}{24}{54} & $0.0969$ & $0.0426$ & $0.0462$ & $0.0532$ & $0.0578$ \\
\Element{Mn}{25}{55} & $0.104$ & $0.0457$ & $0.0498$ & $0.0548$ & $0.0597$ \\
\Element{Fe}{26}{54} & $0.113$ & $0.0497$ & $0.0543$ & $0.0536$ & $0.0585$ \\
\Element{Fe}{26}{56} & $0.11$ & $0.0483$ & $0.0529$ & $0.0558$ & $0.061$ \\
\Element{Fe}{26}{57} & $0.109$ & $0.0478$ & $0.0524$ & $0.057$ & $0.0625$ \\
\Element{Co}{27}{59} & $0.115$ & $0.0503$ & $0.0554$ & $0.0597$ & $0.0656$ \\
\Element{Ni}{28}{58} & $0.124$ & $0.0544$ & $0.06$ & $0.0583$ & $0.0643$ \\
\Element{Ni}{28}{60} & $0.122$ & $0.0531$ & $0.0587$ & $0.0607$ & $0.0671$ \\
\Element{Ni}{28}{61} & $0.121$ & $0.0527$ & $0.0583$ & $0.0621$ & $0.0687$ \\
\Element{Ni}{28}{62} & $0.12$ & $0.0521$ & $0.0576$ & $0.0633$ & $0.07$ \\
\Element{Cu}{29}{63} & $0.124$ & $0.0541$ & $0.0601$ & $0.0634$ & $0.0705$ \\
\Element{Cu}{29}{65} & $0.123$ & $0.0534$ & $0.0594$ & $0.0662$ & $0.0737$ \\
\Element{Zn}{30}{64} & $0.129$ & $0.0558$ & $0.0624$ & $0.0633$ & $0.0707$ \\
\Element{Zn}{30}{66} & $0.127$ & $0.055$ & $0.0615$ & $0.066$ & $0.0739$ \\
\Element{Zn}{30}{67} & $0.127$ & $0.0549$ & $0.0614$ & $0.0677$ & $0.0757$ \\
\Element{Zn}{30}{68} & $0.126$ & $0.0544$ & $0.0609$ & $0.0689$ & $0.0771$ \\
\Element{Ga}{31}{69} & $0.131$ & $0.0565$ & $0.0635$ & $0.0692$ & $0.0779$ \\
\Element{Ga}{31}{71} & $0.13$ & $0.0559$ & $0.0629$ & $0.0721$ & $0.0812$ \\
\Element{Ge}{32}{70} & $0.135$ & $0.0579$ & $0.0655$ & $0.0688$ & $0.0778$ \\
\Element{Ge}{32}{72} & $0.133$ & $0.0572$ & $0.0648$ & $0.0716$ & $0.081$ \\
\Element{Ge}{32}{73} & $0.133$ & $0.057$ & $0.0646$ & $0.073$ & $0.0827$ \\
\Element{Ge}{32}{74} & $0.132$ & $0.0565$ & $0.0641$ & $0.0742$ & $0.0841$ \\
\Element{Ge}{32}{76} & $0.131$ & $0.0563$ & $0.0638$ & $0.0773$ & $0.0877$ \\
\Element{As}{33}{75} & $0.137$ & $0.0588$ & $0.0669$ & $0.0748$ & $0.0851$ \\
\Element{Se}{34}{76} & $0.141$ & $0.06$ & $0.0687$ & $0.0741$ & $0.0849$ \\
\Element{Se}{34}{77} & $0.141$ & $0.06$ & $0.0687$ & $0.0759$ & $0.0869$ \\
\Element{Se}{34}{78} & $0.141$ & $0.0599$ & $0.0686$ & $0.0776$ & $0.0888$ \\
\Element{Se}{34}{80} & $0.141$ & $0.06$ & $0.0687$ & $0.0811$ & $0.0929$ \\
\Element{Se}{34}{82} & $0.141$ & $0.06$ & $0.0687$ & $0.0846$ & $0.0969$ \\
\Element{Br}{35}{79} & $0.146$ & $0.062$ & $0.0713$ & $0.0779$ & $0.0897$ \\
\Element{Br}{35}{81} & $0.146$ & $0.0621$ & $0.0715$ & $0.0817$ & $0.094$ \\
\Element{Kr}{36}{80} & $0.15$ & $0.0634$ & $0.0734$ & $0.0775$ & $0.0897$ \\
\Element{Kr}{36}{82} & $0.15$ & $0.0636$ & $0.0736$ & $0.0813$ & $0.0941$ \\
\Element{Kr}{36}{83} & $0.151$ & $0.0639$ & $0.0739$ & $0.0834$ & $0.0965$ \\
\Element{Kr}{36}{84} & $0.15$ & $0.0638$ & $0.0738$ & $0.0851$ & $0.0984$ \\
\Element{Kr}{36}{86} & $0.151$ & $0.0641$ & $0.0741$ & $0.089$ & $0.103$ \\
\Element{Rb}{37}{85} & $0.156$ & $0.0661$ & $0.0768$ & $0.0858$ & $0.0996$ \\
\Element{Rb}{37}{87} & $0.157$ & $0.0664$ & $0.0771$ & $0.0897$ & $0.104$ \\
\Element{Sr}{38}{86} & $0.16$ & $0.0677$ & $0.0791$ & $0.0855$ & $0.0999$ \\
\Element{Sr}{38}{87} & $0.161$ & $0.068$ & $0.0794$ & $0.0877$ & $0.102$ \\
\Element{Sr}{38}{88} & $0.161$ & $0.0681$ & $0.0795$ & $0.0896$ & $0.105$ \\
\Element{Y}{39}{89} & $0.166$ & $0.07$ & $0.0822$ & $0.0898$ & $0.105$ \\
\Element{Zr}{40}{90} & $0.17$ & $0.0715$ & $0.0843$ & $0.0894$ & $0.105$ \\
\Element{Zr}{40}{91} & $0.168$ & $0.0706$ & $0.0834$ & $0.0901$ & $0.106$ \\
\Element{Zr}{40}{92} & $0.166$ & $0.0694$ & $0.082$ & $0.0902$ & $0.107$ \\
\Element{Zr}{40}{94} & $0.163$ & $0.0679$ & $0.0804$ & $0.0917$ & $0.109$ \\
\Element{Zr}{40}{96} & $0.16$ & $0.0668$ & $0.0792$ & $0.0935$ & $0.111$ \\
\Element{Nb}{41}{93} & $0.171$ & $0.0712$ & $0.0846$ & $0.0903$ & $0.107$ \\
\Element{Mo}{42}{92} & $0.179$ & $0.0746$ & $0.0889$ & $0.0888$ & $0.106$ \\
\Element{Mo}{42}{94} & $0.174$ & $0.0723$ & $0.0864$ & $0.0895$ & $0.107$ \\
\Element{Mo}{42}{95} & $0.173$ & $0.0717$ & $0.0857$ & $0.0904$ & $0.108$ \\
\Element{Mo}{42}{96} & $0.17$ & $0.0703$ & $0.0842$ & $0.0904$ & $0.108$ \\
\Element{Mo}{42}{97} & $0.169$ & $0.0701$ & $0.084$ & $0.0919$ & $0.11$ \\
\Element{Mo}{42}{98} & $0.167$ & $0.0689$ & $0.0826$ & $0.0918$ & $0.11$ \\
\Element{Mo}{42}{100} & $0.162$ & $0.0667$ & $0.0802$ & $0.0921$ & $0.111$ \\
\Element{Ru}{44}{96} & $0.182$ & $0.0754$ & $0.091$ & $0.0891$ & $0.108$ \\
\Element{Ru}{44}{98} & $0.178$ & $0.0733$ & $0.0887$ & $0.09$ & $0.109$ \\
\Element{Ru}{44}{99} & $0.177$ & $0.0726$ & $0.088$ & $0.0908$ & $0.11$ \\
\Element{Ru}{44}{100} & $0.174$ & $0.0714$ & $0.0866$ & $0.0909$ & $0.11$ \\
\Element{Ru}{44}{101} & $0.173$ & $0.0709$ & $0.0861$ & $0.0919$ & $0.112$ \\
\Element{Ru}{44}{102} & $0.17$ & $0.0697$ & $0.0847$ & $0.0918$ & $0.112$ \\
\Element{Ru}{44}{104} & $0.166$ & $0.0679$ & $0.0827$ & $0.0926$ & $0.113$ \\
\Element{Rh}{45}{103} & $0.175$ & $0.0713$ & $0.0871$ & $0.0919$ & $0.112$ \\
\Element{Pd}{46}{102} & $0.183$ & $0.0746$ & $0.0914$ & $0.0908$ & $0.111$ \\
\Element{Pd}{46}{104} & $0.179$ & $0.0729$ & $0.0895$ & $0.092$ & $0.113$ \\
\Element{Pd}{46}{105} & $0.178$ & $0.0725$ & $0.089$ & $0.0929$ & $0.114$ \\
\Element{Pd}{46}{106} & $0.176$ & $0.0714$ & $0.0878$ & $0.0931$ & $0.114$ \\
\Element{Pd}{46}{108} & $0.172$ & $0.0698$ & $0.086$ & $0.094$ & $0.116$ \\
\Element{Pd}{46}{110} & $0.169$ & $0.0684$ & $0.0845$ & $0.0951$ & $0.118$ \\
\Element{Ag}{47}{107} & $0.18$ & $0.0729$ & $0.0901$ & $0.093$ & $0.115$ \\
\Element{Ag}{47}{109} & $0.177$ & $0.0716$ & $0.0887$ & $0.0945$ & $0.117$ \\
\Element{Cd}{48}{106} & $0.187$ & $0.0757$ & $0.0939$ & $0.0915$ & $0.114$ \\
\Element{Cd}{48}{110} & $0.181$ & $0.0731$ & $0.091$ & $0.0944$ & $0.118$ \\
\Element{Cd}{48}{111} & $0.18$ & $0.0726$ & $0.0904$ & $0.0952$ & $0.119$ \\
\Element{Cd}{48}{112} & $0.179$ & $0.0719$ & $0.0896$ & $0.0959$ & $0.12$ \\
\Element{Cd}{48}{113} & $0.178$ & $0.0714$ & $0.0891$ & $0.0967$ & $0.121$ \\
\Element{Cd}{48}{114} & $0.177$ & $0.0709$ & $0.0886$ & $0.0975$ & $0.122$ \\
\Element{Cd}{48}{116} & $0.175$ & $0.0702$ & $0.0877$ & $0.0994$ & $0.124$ \\
\Element{In}{49}{113} & $0.184$ & $0.0737$ & $0.0923$ & $0.0963$ & $0.121$ \\
\Element{In}{49}{115} & $0.182$ & $0.0727$ & $0.0912$ & $0.0979$ & $0.123$ \\
\Element{Sn}{50}{116} & $0.186$ & $0.0743$ & $0.0936$ & $0.098$ & $0.124$ \\
\Element{Sn}{50}{117} & $0.185$ & $0.0739$ & $0.0932$ & $0.099$ & $0.125$ \\
\Element{Sn}{50}{118} & $0.184$ & $0.0733$ & $0.0925$ & $0.0996$ & $0.126$ \\
\Element{Sn}{50}{119} & $0.183$ & $0.073$ & $0.0921$ & $0.101$ & $0.127$ \\
\Element{Sn}{50}{120} & $0.182$ & $0.0724$ & $0.0915$ & $0.101$ & $0.128$ \\
\Element{Sn}{50}{122} & $0.18$ & $0.0716$ & $0.0906$ & $0.103$ & $0.13$ \\
\Element{Sn}{50}{124} & $0.179$ & $0.0709$ & $0.0898$ & $0.105$ & $0.133$ \\
\Element{Sb}{51}{121} & $0.183$ & $0.0725$ & $0.0923$ & $0.0996$ & $0.127$ \\
\Element{Sb}{51}{123} & $0.182$ & $0.072$ & $0.0917$ & $0.102$ & $0.129$ \\
\Element{Te}{52}{122} & $0.184$ & $0.0725$ & $0.0929$ & $0.0976$ & $0.125$ \\
\Element{Te}{52}{124} & $0.183$ & $0.0719$ & $0.0922$ & $0.0995$ & $0.128$ \\
\Element{Te}{52}{125} & $0.182$ & $0.0717$ & $0.092$ & $0.101$ & $0.129$ \\
\Element{Te}{52}{126} & $0.182$ & $0.0713$ & $0.0915$ & $0.101$ & $0.13$ \\
\Element{Te}{52}{128} & $0.18$ & $0.0707$ & $0.0909$ & $0.103$ & $0.133$ \\
\Element{Te}{52}{130} & $0.179$ & $0.0702$ & $0.0903$ & $0.105$ & $0.135$ \\
\Element{I}{53}{127} & $0.183$ & $0.0716$ & $0.0925$ & $0.1$ & $0.129$ \\
\Element{Xe}{54}{128} & $0.184$ & $0.0715$ & $0.0931$ & $0.098$ & $0.128$ \\
\Element{Xe}{54}{129} & $0.184$ & $0.0715$ & $0.0931$ & $0.0994$ & $0.129$ \\
\Element{Xe}{54}{130} & $0.183$ & $0.0712$ & $0.0927$ & $0.1$ & $0.131$ \\
\Element{Xe}{54}{131} & $0.184$ & $0.0713$ & $0.0928$ & $0.102$ & $0.132$ \\
\Element{Xe}{54}{132} & $0.183$ & $0.0709$ & $0.0924$ & $0.102$ & $0.133$ \\
\Element{Xe}{54}{134} & $0.182$ & $0.0706$ & $0.0921$ & $0.105$ & $0.136$ \\
\Element{Xe}{54}{136} & $0.181$ & $0.0702$ & $0.0915$ & $0.107$ & $0.139$ \\
\Element{Cs}{55}{133} & $0.185$ & $0.0715$ & $0.0937$ & $0.101$ & $0.133$ \\
\Element{Ba}{56}{134} & $0.185$ & $0.0712$ & $0.094$ & $0.0992$ & $0.131$ \\
\Element{Ba}{56}{135} & $0.186$ & $0.0714$ & $0.0943$ & $0.101$ & $0.133$ \\
\Element{Ba}{56}{136} & $0.185$ & $0.0711$ & $0.0939$ & $0.102$ & $0.134$ \\
\Element{Ba}{56}{137} & $0.185$ & $0.0713$ & $0.0941$ & $0.103$ & $0.136$ \\
\Element{Ba}{56}{138} & $0.184$ & $0.0708$ & $0.0936$ & $0.104$ & $0.137$ \\
\Element{La}{57}{139} & $0.187$ & $0.0713$ & $0.0948$ & $0.103$ & $0.136$ \\
\Element{Ce}{58}{140} & $0.188$ & $0.0713$ & $0.0955$ & $0.101$ & $0.135$ \\
\Element{Ce}{58}{142} & $0.183$ & $0.0691$ & $0.0929$ & $0.1$ & $0.135$ \\
\Element{Pr}{59}{141} & $0.19$ & $0.0718$ & $0.0968$ & $0.0998$ & $0.135$ \\
\Element{Nd}{60}{142} & $0.191$ & $0.0719$ & $0.0976$ & $0.0983$ & $0.133$ \\
\Element{Nd}{60}{143} & $0.189$ & $0.0709$ & $0.0964$ & $0.098$ & $0.133$ \\
\Element{Nd}{60}{144} & $0.186$ & $0.0696$ & $0.0949$ & $0.0974$ & $0.133$ \\
\Element{Nd}{60}{145} & $0.184$ & $0.0687$ & $0.0938$ & $0.0974$ & $0.133$ \\
\Element{Nd}{60}{146} & $0.181$ & $0.0675$ & $0.0924$ & $0.0968$ & $0.132$ \\
\Element{Nd}{60}{148} & $0.176$ & $0.0652$ & $0.0897$ & $0.0957$ & $0.132$ \\
\Element{Nd}{60}{150} & $0.17$ & $0.0623$ & $0.0863$ & $0.0935$ & $0.129$ \\
\Element{Sm}{62}{144} & $0.193$ & $0.0719$ & $0.099$ & $0.0951$ & $0.131$ \\
\Element{Sm}{62}{147} & $0.187$ & $0.069$ & $0.0955$ & $0.0946$ & $0.131$ \\
\Element{Sm}{62}{148} & $0.184$ & $0.0678$ & $0.0941$ & $0.0941$ & $0.131$ \\
\Element{Sm}{62}{149} & $0.182$ & $0.0671$ & $0.0933$ & $0.0942$ & $0.131$ \\
\Element{Sm}{62}{150} & $0.178$ & $0.0652$ & $0.091$ & $0.0926$ & $0.129$ \\
\Element{Sm}{62}{152} & $0.171$ & $0.062$ & $0.0872$ & $0.09$ & $0.127$ \\
\Element{Sm}{62}{154} & $0.167$ & $0.0603$ & $0.0852$ & $0.0895$ & $0.126$ \\
\Element{Eu}{63}{151} & $0.18$ & $0.0656$ & $0.0921$ & $0.0916$ & $0.129$ \\
\Element{Eu}{63}{153} & $0.17$ & $0.0611$ & $0.0868$ & $0.0873$ & $0.124$ \\
\Element{Gd}{64}{154} & $0.172$ & $0.0616$ & $0.088$ & $0.0866$ & $0.124$ \\
\Element{Gd}{64}{155} & $0.17$ & $0.0609$ & $0.0872$ & $0.0866$ & $0.124$ \\
\Element{Gd}{64}{156} & $0.169$ & $0.0601$ & $0.0863$ & $0.0864$ & $0.124$ \\
\Element{Gd}{64}{157} & $0.168$ & $0.0599$ & $0.086$ & $0.0871$ & $0.125$ \\
\Element{Gd}{64}{158} & $0.166$ & $0.0591$ & $0.085$ & $0.0867$ & $0.125$ \\
\Element{Gd}{64}{160} & $0.163$ & $0.0579$ & $0.0835$ & $0.0868$ & $0.125$ \\
\Element{Tb}{65}{159} & $0.187$ & $0.0677$ & $0.096$ & $0.0978$ & $0.139$ \\
\Element{Dy}{66}{160} & $0.167$ & $0.0586$ & $0.0857$ & $0.0834$ & $0.122$ \\
\Element{Dy}{66}{161} & $0.167$ & $0.0585$ & $0.0856$ & $0.0842$ & $0.123$ \\
\Element{Dy}{66}{162} & $0.165$ & $0.0577$ & $0.0846$ & $0.0839$ & $0.123$ \\
\Element{Dy}{66}{163} & $0.164$ & $0.0575$ & $0.0844$ & $0.0845$ & $0.124$ \\
\Element{Dy}{66}{164} & $0.162$ & $0.0566$ & $0.0833$ & $0.0841$ & $0.124$ \\
\Element{Ho}{67}{165} & $0.169$ & $0.0592$ & $0.0871$ & $0.0865$ & $0.127$ \\
\Element{Er}{68}{164} & $0.167$ & $0.0575$ & $0.0857$ & $0.0812$ & $0.121$ \\
\Element{Er}{68}{166} & $0.164$ & $0.0566$ & $0.0846$ & $0.0815$ & $0.122$ \\
\Element{Er}{68}{167} & $0.164$ & $0.0563$ & $0.0842$ & $0.0819$ & $0.123$ \\
\Element{Er}{68}{168} & $0.162$ & $0.0556$ & $0.0834$ & $0.0818$ & $0.123$ \\
\Element{Er}{68}{170} & $0.16$ & $0.0546$ & $0.0821$ & $0.0819$ & $0.123$ \\
\Element{Tm}{69}{169} & $0.172$ & $0.0596$ & $0.0889$ & $0.0863$ & $0.129$ \\
\Element{Yb}{70}{170} & $0.165$ & $0.0561$ & $0.0853$ & $0.0801$ & $0.122$ \\
\Element{Yb}{70}{171} & $0.164$ & $0.0557$ & $0.0848$ & $0.0804$ & $0.122$ \\
\Element{Yb}{70}{172} & $0.163$ & $0.055$ & $0.084$ & $0.0802$ & $0.122$ \\
\Element{Yb}{70}{173} & $0.162$ & $0.0547$ & $0.0835$ & $0.0804$ & $0.123$ \\
\Element{Yb}{70}{174} & $0.161$ & $0.0542$ & $0.083$ & $0.0805$ & $0.123$ \\
\Element{Yb}{70}{176} & $0.159$ & $0.0534$ & $0.082$ & $0.0809$ & $0.124$ \\
\Element{Lu}{71}{175} & $0.154$ & $0.0509$ & $0.0794$ & $0.0745$ & $0.116$ \\
\Element{Lu}{71}{176} & $0.153$ & $0.0506$ & $0.0791$ & $0.0748$ & $0.117$ \\
\Element{Hf}{72}{176} & $0.164$ & $0.0548$ & $0.085$ & $0.0791$ & $0.123$ \\
\Element{Hf}{72}{177} & $0.164$ & $0.0546$ & $0.0848$ & $0.0796$ & $0.124$ \\
\Element{Hf}{72}{178} & $0.163$ & $0.0541$ & $0.0842$ & $0.0797$ & $0.124$ \\
\Element{Hf}{72}{179} & $0.162$ & $0.0538$ & $0.0838$ & $0.08$ & $0.125$ \\
\Element{Hf}{72}{180} & $0.161$ & $0.0534$ & $0.0833$ & $0.0801$ & $0.125$ \\
\Element{Ta}{73}{181} & $0.163$ & $0.054$ & $0.0847$ & $0.0799$ & $0.125$ \\
\Element{W}{74}{182} & $0.166$ & $0.0545$ & $0.086$ & $0.0795$ & $0.125$ \\
\Element{W}{74}{183} & $0.165$ & $0.0541$ & $0.0855$ & $0.0797$ & $0.126$ \\
\Element{W}{74}{184} & $0.164$ & $0.0537$ & $0.085$ & $0.0799$ & $0.126$ \\
\Element{W}{74}{186} & $0.162$ & $0.0531$ & $0.0842$ & $0.0803$ & $0.127$ \\
\Element{Re}{75}{185} & $0.168$ & $0.0551$ & $0.0874$ & $0.0808$ & $0.128$ \\
\Element{Re}{75}{187} & $0.166$ & $0.0543$ & $0.0864$ & $0.0811$ & $0.129$ \\
\Element{Os}{76}{186} & $0.165$ & $0.0535$ & $0.0861$ & $0.0774$ & $0.125$ \\
\Element{Os}{76}{187} & $0.165$ & $0.0533$ & $0.0858$ & $0.0779$ & $0.125$ \\
\Element{Os}{76}{188} & $0.164$ & $0.0529$ & $0.0853$ & $0.0779$ & $0.126$ \\
\Element{Os}{76}{189} & $0.163$ & $0.0527$ & $0.085$ & $0.0783$ & $0.126$ \\
\Element{Os}{76}{190} & $0.163$ & $0.0523$ & $0.0846$ & $0.0785$ & $0.127$ \\
\Element{Os}{76}{192} & $0.161$ & $0.0519$ & $0.084$ & $0.0791$ & $0.128$ \\
\Element{Ir}{77}{191} & $0.167$ & $0.0539$ & $0.0872$ & $0.0797$ & $0.129$ \\
\Element{Ir}{77}{193} & $0.166$ & $0.0534$ & $0.0866$ & $0.0804$ & $0.13$ \\
\Element{Pt}{78}{194} & $0.165$ & $0.0526$ & $0.0863$ & $0.0782$ & $0.128$ \\
\Element{Pt}{78}{195} & $0.165$ & $0.0523$ & $0.0859$ & $0.0785$ & $0.129$ \\
\Element{Pt}{78}{196} & $0.164$ & $0.052$ & $0.0856$ & $0.0787$ & $0.129$ \\
\Element{Pt}{78}{198} & $0.163$ & $0.0514$ & $0.0848$ & $0.0792$ & $0.13$ \\
\Element{Au}{79}{197} & $0.166$ & $0.0523$ & $0.0866$ & $0.0781$ & $0.129$ \\
\Element{Hg}{80}{198} & $0.167$ & $0.0523$ & $0.0873$ & $0.0772$ & $0.129$ \\
\Element{Hg}{80}{199} & $0.167$ & $0.0522$ & $0.0872$ & $0.0777$ & $0.13$ \\
\Element{Hg}{80}{200} & $0.165$ & $0.0516$ & $0.0864$ & $0.0775$ & $0.13$ \\
\Element{Hg}{80}{201} & $0.165$ & $0.0514$ & $0.0861$ & $0.0778$ & $0.13$ \\
\Element{Hg}{80}{202} & $0.163$ & $0.0509$ & $0.0855$ & $0.0776$ & $0.13$ \\
\Element{Hg}{80}{204} & $0.161$ & $0.0502$ & $0.0845$ & $0.0777$ & $0.131$ \\
\Element{Tl}{81}{203} & $0.166$ & $0.0515$ & $0.0869$ & $0.0775$ & $0.131$ \\
\Element{Tl}{81}{205} & $0.164$ & $0.0507$ & $0.0859$ & $0.0777$ & $0.132$ \\
\Element{Pb}{82}{204} & $0.166$ & $0.0511$ & $0.0871$ & $0.076$ & $0.13$ \\
\Element{Pb}{82}{206} & $0.164$ & $0.0503$ & $0.0861$ & $0.0761$ & $0.13$ \\
\Element{Pb}{82}{207} & $0.163$ & $0.05$ & $0.0857$ & $0.0762$ & $0.131$ \\
\Element{Pb}{82}{208} & $0.162$ & $0.0495$ & $0.085$ & $0.076$ & $0.131$ \\
\Element{Bi}{83}{209} & $0.161$ & $0.0486$ & $0.0845$ & $0.0738$ & $0.128$ \\
\Element{Th}{90}{232} & $0.129$ & $0.033$ & $0.0683$ & $0.0521$ & $0.108$ \\
\Element{U}{92}{238} & $0.12$ & $0.0289$ & $0.0639$ & $0.0459$ & $0.101$ \\\hline
\end{supertabular}

\bibliographystyle{utcaps_mod}
\bibliography{BIB}

\end{document}